\documentclass[preprint,12pt]{elsarticle}
\newtheorem{theorem}{Theorem}
\newtheorem{definition}{Definition}

\usepackage{amssymb}
\usepackage[ruled,vlined]{algorithm2e}
\usepackage{graphicx}
\usepackage{subfigure}
\usepackage{booktabs}
\usepackage{url}
\usepackage{multirow}
\usepackage{array}
 
\journal{Physica A}

\begin{document}

\begin{frontmatter}



\title{Correlation Analysis of Nodes Identifies Real Communities in Networks}


\author[1]{Jingming Zhang}
\ead{zhangjm17@lzu.edu.cn}
\author[1]{Jianjun Cheng}
\ead{chengjianjun@lzu.edu.cn}
\author[1]{Xing Su}
\author[1]{Xinghong Yin}
\author[1]{Shiyan Zhao}
\author[1]{Xiaoyun Chen\corref{cor}}
\ead{chenxy@lzu.edu.cn}
\cortext[cor]{Corresponding author.}
\address[1]{School of Information Science and Engineering, Lanzhou University, South Tianshui Road No. 222, Lanzhou 730000, P. R. China }

\begin{abstract}
A significant problem in analysis of complex network is to reveal community structure, in which network nodes are tightly connected in the same communities, between which there are sparse connections. Previous algorithms for community detection in real-world networks have the shortcomings of high complexity or requiring for prior information such as the number or sizes of communities or are unable to obtain the same resulting partition in multiple runs. In this paper, we proposed a simple and effective algorithm that uses the correlation of nodes alone, which requires neither optimization of predefined objective function nor information about the number or sizes of communities. We test our algorithm on real-world and synthetic graphs whose community structure is already known and observe that the proposed algorithm detects this known structure with high applicability and reliability. We also apply the algorithm to some networks whose community structure is unknown and find that it detects deterministic and informative community partitions in these cases.
\end{abstract}


\end{frontmatter}


\section{Introduction}
Community detection in networks has been intensively investigated in recent years, which has significant contribution to our understanding of complex systems\cite{FORTUNATO20161,newman2012communities}. Community structure is one of the most important features for complex networks, in which nodes are tightly connected in the same communities, between which there are only sparse connections\cite{girvan2002community}. More importantly, community structures are often connected with functional and organizational characteristics of latent networks\cite{PhysRevE.69.026113,PhysRevE.90.062805}. Therefore community detection has much practical significance.

Many community detection methods have been proposed in recent years. These methods attempt to disclose community structure characteristics in networks from various perspectives. In order to minimize the inter-group edges, the traditional graph partitioning methods divide the nodes into a predefined number of groups with definite size. Hierarchical clustering techniques uncover the grouping structure of a graph, either in divisive style or agglomerative way\cite{newman2012communities}. Spectral clustering algorithms partition a network into groups by using the eigenvectors of affinity matrices\cite{PhysRevE.74.036104,Newman8577}. Modularity maximization methods convert the problem of community detection into an task of optimizing the modularity function to get the optimal community partitioning\cite{PhysRevE.70.066111}. Most of the above methods can be regarded as global approaches, and they suffer from common limitations that require for a priori knowledge such as the number or sizes of communities, which are usually uncertain and unavailable in advance. Moreover, many global methods tend to be computationally demanding in spite of high accuracy. Therefore, it is non-trivial to balance between accuracy and efficiency for community detection.

Many local approaches have been proposed to solve the limitations mentioned above. They depend only on local information of nodes. Local methods are empirically classified into four categories: label propagation based algorithms, density-based clustering methods, dynamic-based approaches and local expansion optimization methods. Usha Nandini Raghavan et al. proposed the first label propagation based algorithm in 2007\cite{PhysRevE.76.036106}. Moreover, density-based clustering methods are also noteworthy. Tao You et al.\cite{YOU2016221} proposed a density-based clustering algorithm combined IsoMap and Fdp technique. However, this method need to supply appropriate parameters like other density-based clustering methods\cite{GONG201471,Xu:2007:SSC:1281192.1281280}. Notably, dynamic-based approaches have employed varios techniques such as random walk and particle system. Walktrap\cite{pons2005computing} algorithm is based on random walk. Quiles et al. proposed a novel community detection algorithm using particle system. Furthermore, Attractor\cite{Shao2015Community} method based on distance dynamics is introduced in 2015. Unfortunately, these methods accompany some drawbacks such as randomness and weak robustness. with respect to local expansion optimization methods, they are usually used for local community detection in large networks because of the advantages both in accuracy and efficiency. However, they are also sensitive to initial seeds and built-in parameters. In 2012, Kun et al. has proposed an efficient algorithm based on the strength of weak ties hypothesis\cite{li2012efficient}. Tao et al.\cite{WANG20181344} proposed a local community detection method based on local similarity and degree clustering information. Krista introduced a novel community detection algorithm using maximal neighbor similarity\cite{vzalik2015maximal}. But these methods produce unstable or relatively bad results.    

In this paper, we present a novel method for community detection which is termed as CAN(Correlation Analysis of Nodes). It is inspired by a simple idea that a pair of nodes in the same community are more correlated than between different communities. Compared with previous state-of-the-art algorithms that mentioned above, the proposed method dose not acquire prior knowledge. Moreover, it avoids instability of some methods.

The remainder of this paper is organized as follows. The proposed algorithm is described in detail in section \ref{CANag}. Section \ref{ED} presents experimental results and discusses the selection of parameters. And the paper is ended with a conclusion in section \ref{conclusion}.

\section{CAN algorithm}\label{CANag}
In this section, we propose a community detection algorithm motivated by a simple idea that nodes in the same community are more correlated than between different communities.	A network can be represented as a graph $G(V,E)$ with a set of vertices $V = (v_1,v_2,...,v_n)$ that represents objects and a set of edges $E$ that models relationships between each pair of nodes that interact with each other. Communities are defined as intrinsic groups of densely interconnected nodes that have sparse connection with the other network. On the basis of this definition, the nodes in the same community tend to have the same feature such as link pattern. Link pattern of node $v\in V$ is an attribute vector constructed from the similarity of node $v$ to the other nodes, which defined as eq.\ref{LinkPattern}. In order to measure the correlation between a pair of nodes, we firstly construct link pattern of every node using eq.\ref{simlarity}. Secondly, we find all pairs of nodes with strong correlation in the networks using Pearson correlation coefficient. And then we can get initial partition on the basis of second step, which can be seen from subsection \ref{GIP}. Of course, we finally identify intrinsic and definite community structures by adjusting fuzzy nodes and some small-scale communities. To understand our approach explicitly, we first have a big picture with our algorithm outlined in Algorithm\ref{algorithm}, and then expound every step in our method at length. 
\begin{algorithm}\label{algorithm}	
	\DontPrintSemicolon
	\SetAlgoNoLine  
	\KwIn{A network $G=(V,E)$ and the threshold Correlation Coefficient $\beta$.}
	\KwOut{Community structure of network $G$.}
	\For{ $v \in V$}{ 
		\For{$u \in V$}{
			Calculate Similarity Matrix using equation \ref{simlarity}.\\
		}
	}
	$StrongCorrrelation \leftarrow \varnothing$\\
	\For{ $v \in V$}{ 
		\For{$u \in V$}{
		 	\If{$r_{u,v}>\beta$}{
		 		$StrongCorrelation  \leftarrow StrongCorrelation + (u,v)$\\
		 	}
		}
	}
	Get initial partition $P$ using Theorem \ref{theorem1}\\
	$FuzzyNode \leftarrow V - P$\\
	\For{$v \in FuzzyNode$}{
		Merge node $v$ into corresponding partition using equation \ref{fuzzynode}\\
	}
	\For{$p \in P$}{
		\If{$p < \lambda$}{
			\For{$v \in p$}{
				Merge node $v$ into corresponding partition using equation \ref{fuzzynode}\\
			}
		}
	}
	\caption{CAN algorithm\label{CAN}}
\end{algorithm}

\subsection{Get Initial Partition}\label{GIP}
In order to get initial partition, we firstly need to calculate the link pattern of every node. The link pattern of node $v_i\in V$ is defined as
\begin{equation}
	\label{LinkPattern}
	LP(v_i) = [Sim(v_i,v_1),Sim(v_i,v_2),...,Sim(v_i,v_k),...,Sim(v_i,v_n)]
\end{equation}
where $Sim(v_i,v_1)$ represents the similarity between nodes $v_i\in V$ and $v_1\in V$, and $n$ is the number of nodes in network.
Many node similarity metrics based on local information have been proposed \cite{zhou2009predicting}. Salton's Cosine index\cite{Salton:1986:IMI:576628} is selected as the node similarity measure in our method. Salton's Cosine is defined as:
\begin{equation}\label{simlarity}
Sim(v_i,v_j)=\frac{\mid \Gamma_{v_iv_j} \mid}{\sqrt{\mid \Gamma_{v_i}\mid \times \mid \Gamma_{v_j}\mid}}
\end{equation} 
where $\Gamma_{v_i}$ is the neighbor set of node $v_i$, i.e., $\Gamma_{v_i}=\{v_j|(v_j,v_i) \in E\}\cup \{v_i\}$, $\Gamma_{v_iv_j}$ is the number of common neighbors of nodes $v_i$ and $v_j$, i.e., $\Gamma_{v_iv_j}=\Gamma_{v_i}\cap\Gamma_{v_j}$ . 

Secondly, we can analyze the correlation between nodes $u \in V$ and $v \in V$ using link pattern and  Pearson correlation coefficient \cite{Pearson1895Proceedings}. This is
\begin{equation}\label{pcc}
r_{LP(u),LP(v)}=\frac{\sum_{i=1}^{n}(x_i-\bar{X})(y_i-\bar{Y})}{n\sigma_X \sigma_Y}=\frac{\sum_{i=1}^{n}(x_iy_i) - n\bar{X} \bar{Y}}{n\sigma_X \sigma_Y}
\end{equation}
where n is the number of nodes, $X$ and $Y$ are the respective link pattern of nodes u and v, $x_i$ and $y_i$ serve as the values of $Sim(u,i)$ and $Sim(v,i)$ respectively, $\bar{X}$ and $\bar{Y}$ are the respective mean values of $X$ and $Y$,
$\sigma_X$ and $\sigma_Y$ represent the standard deviations of $X$ and $Y$ separately, and $\sum(x_iy_i)$ is the sum of the $XY$ cross-product. For purpose of finding all pairs of nodes with strong correlation, we put forward the following definition and theorem on the strength of Pearson correlation coefficient in complex networks.
\begin{definition}
	Given a correlation threshold $\beta$, if $r_{LP(v_i),LP(v_j)} > \beta$, then nodes $v_i$ and $v_j$ are strong correlation , and both nodes are in the same community. 
\end{definition}
Obviously,  "strong correlation" is an equivalence relation:
\begin{enumerate}[(1)]
	\item symmetric: If $(v_i,v_j)$ is strong correlation, then $(v_j,v_i)$ is also strong correlation.
	\item transitive: If $(v_i,v_k)$ is strong correlation and $(v_k,v_j)$ is strong correlation, then $(v_i,v_j)$ is strong correlation as well.
	\item reflexive: $(v_i,v_i)$ is strong correlation.
\end{enumerate}
\begin{theorem}\label{theorem1}
	If $r_{LP(v_i),LP(v_j)} > \beta$ and $r_{LP(v_i),LP(v_k)} > \beta$, then $(v_j,v_k)$ is strong correlation in the network, and $\{v_i,v_j,v_k\}$ are in the same community.
\end{theorem}
Proof of theorem \ref{theorem1} is very easy on the basis of the properties of the equivalence relation. Theorem \ref{theorem1} can be used to guide the process of community detection. In fact, a community can be regard as an equivalence class. Consequently, we need only to find all equivalence classes in the graph.

\begin{figure}[!htb]
	\centering
	\makebox[\textwidth][c]{
		\includegraphics[width=0.99\textwidth]{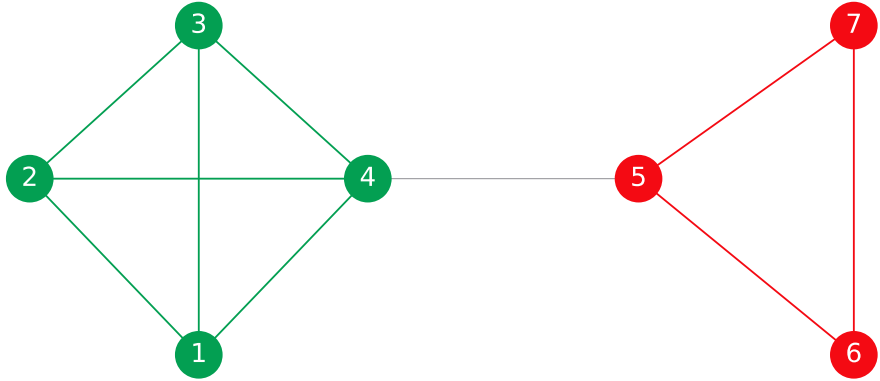} }
	\caption{An example network with community structure.}\label{fig:example}
\end{figure}

In order to show more details about how to get initial partition, an example network with 7 nodes and 10 edges is shown in Fig. \ref{fig:example}.

For the node $v_i(i\in [1,...,7])$ in the network shown in Fig.\ref{fig:example}, the link pattern of each node can be obtained using eq.\ref{LinkPattern} and eq.\ref{simlarity}. The link pattern of each node are shown in Table \ref{tab:example}.
\begin{table}[htbp] 
	\centering
	\caption{The Link pattern of each node}
	\label{tab:example}
	\begin{tabular}{lccccccr}
		\hline
		\textbf{Link Pattern} &  &  &  & & & & \\
		\hline
		$LP(v_1)$ & [1.000     &  1.000   & 1.000 & 0.894 & 0.250 & 0.000 & 0.000]\\
		$LP(v_2)$ & [1.000     &  1.000   & 1.000 & 0.894 & 0.250 & 0.000 & 0.000]\\
		$LP(v_3)$ & [1.000     &  1.000   & 1.000 & 0.894 & 0.250 & 0.000 & 0.000]\\
		$LP(v_4)$ & [0.894     &  0.894   & 0.894 & 1.000 & 0.447 & 0.258 & 0.258]\\
		$LP(v_5)$ & [0.250     &  0.250   & 0.250 & 0.447 & 1.000 & 0.866 & 0.866]\\
		$LP(v_6)$ & [0.000     &  0.000   & 0.000 & 0.258 & 0.866 & 1.000 & 1.000]\\
		$LP(v_7)$ & [0.000     &  0.000   & 0.000 & 0.258 & 0.866 & 1.000 & 1.000]\\
		\hline
	\end{tabular}
\end{table}	

Then, we can get all pairs of nodes with strong correlation using eq.\ref{pcc} and the link pattern of each node. Given the correlation threshold $\beta=0.8$, we obtain nine pairs of nodes with strong correlation, i.e., $r_{LP(v_1),LP(v_2)}=1.0$, $r_{LP(v_1),LP(v_3)}=1.0$, $r_{LP(v_1),LP(v_4)}=0.98$, $r_{LP(v_2),LP(v_3)}=1.0$, $r_{LP(v_2),LP(v_4)}=0.98$, $r_{LP(v_3),LP(v_4)}=0.98$, $r_{LP(v_5),LP(v_6)}=0.97$, $r_{LP(v_5),LP(v_7)}=0.97$ and $r_{LP(v_6),LP(v_7)}=1.0$. So, the set of strong correlation is
\begin{displaymath}
\{(1,2),(1,3),(1,4),(2,3),(2,4),(3,4),(5,6),(5,7),(6,7)\}
\end{displaymath}

Next, we employ the theorem \ref{theorem1} and above set to get the initial partition. For example, $(1,2)$ and $(1,3)$ are strong correlation, then $(2,3)$ is strong correlation, and nodes $\{1,2,3\}$ are in the same community. Consequently, we finally get the initial partition is $\{\{1,2,3,4\},\{5,6,7\}\}$.

\subsection{Merge fuzzy nodes and small partitions}\label{mfnsp}
In this subsection, we will explain how to merger fuzzy nodes and small partitions using initial partition obtained before. Considerable methods have been proposed to handle the fuzzy nodes, which mean the node that dose not belong to any initial partition in our paper. To merge these fuzzy nodes, we adopt Jaccad index\cite{Jaccard1901Etude} as the similarity measure between node $v$ and partition $C_i(i=1...m)$. The Jaccard index is calculated as the ratio of common neighbors of node $i$ and $j$, normalized by the sum of the neighbors of both nodes:
\begin{equation}
Jaccard(i,j) = \frac{\mid \Gamma_{ij} \mid}{\mid \Gamma_i \cup \Gamma_j\mid}
\end{equation}
Given a fuzzy node $v$, we use eq.\ref{fuzzynode} to determine to which partition it should belong. 
\begin{equation}\label{fuzzynode}
v \in \arg max \sum_{j \in C_i}Jaccad(j,v), \qquad(i=1,...m)
\end{equation}
which means that node $v$ belongs to the community with which it is the most similar.

Each partition that does not satisfy community criteria merges with the other partition between which is the most similar. In our method, community unsatisfied community criteria is defined as the community that the size of community is less than $\lambda$. we can find how set this parameter in section \ref{discussion}. The pseudo-code can be seen in Algorithm \ref{algorithm}.

\subsection{Time complexity analysis}
	Our algorithm consists of calculating similarity matrix, finding pairs of nodes with strong correlation and forming the initial communities, and adjusting the partitions and nodes among communities. Calculation of similarity matrix requires $O(n^2)$ where $n$ is the number of nodes. Choosing all pairs of node with strong correlation costs $O(n\log_2n)$. Then preliminary communities can be formed in $O(\log_2n)$ time complexity. Adjusting partitions and nodes among communities needs $O(vM+sM)$ where $M$ is the number of initial partitions, $v$ is the maximum number of fuzzy nodes and $s$ is the maximal size of community that dose not satisfy community criteria. The total time complexity of the proposed algorithm is $O(n^2+(v+s)M+(n+1)\log_2n)$. For a community structure in a network, $M$ is far smaller than the number of nodes $n$. Therefore, the time complexity of CAN can be simplified as $O(n^2)$.
	 
\section{Experiments and discussion}\label{ED}
In this section, we test the performance of the CAN algorithm on various networks that have widely used in community detection, and compare to some state-of-the-art algorithms[namely, CNM\cite{PhysRevE.70.066111}, LPA\cite{PhysRevE.76.036106}, Isofdp\cite{YOU2016221}, Walktrap\cite{pons2005computing}, Particle\cite{quiles2016dynamical} and Attractor\cite{Shao2015Community} methods] on real-world networks and synthetic networks to illustrate the effectiveness of the proposed method on revealing the community structures. Furthermore, we adopt two critical evaluation criteria, i.e., normalized mutual information and modularity, to evaluate the quality of various community detection algorithm. Thereafter, we will analyze how to set the threshold of correlation coefficient $\beta$, and the parameter $\lambda$ in Section \ref{discussion}.
\subsection{Evalution criteria}
To evaluate the performance of a community detection algorithm in the experiments, we introduce two remarkable evaluation criteria, modularity and normalized mutual information. 
The modularity is an commonly used quality measure proposed by Grivan and Newman\cite{Newman2004Finding}. It is based on the assumption that a community structure is not detected in random graphs. It can be defined as:
\begin{equation}
Q = \sum_{i=1}(e_{ii} - a_i^2)
\end{equation}  
where $a_i=\sum_{j}e_{ij}$ and $e_{ij}$ is the fraction of edges that connect nodes in community i to those in community j, $e_{ii}$ is the fraction of edges that fall within community $i$.

Normalized mutual information(NMI) is an information-theory based measurement, which is extensively used in evaluating the quality of our method on networks with known groud truth community structures\cite{Danon2005Comparing}. It can be calculated as:
\begin{equation}
NMI(X,Y)=\frac{-2\sum_{i=1}^{C_X}\sum_{j=1}^{C_Y}\log{\frac{N_{ij}N}{N_{i.}N_{.j}}}}{\sum_{i=1}^{C_X}N_{i.}\log{\frac{N_{i.}}{N}}+\sum_{j=1}^{C_Y}N_{.j}\log{\frac{N_{.j}}{N}}}
\end{equation}
where $X$ and $Y$ denotes uncovered community partition and real partition respectively; $C_X$, $C_Y$ is the number of communities in $X$ and $Y$; $N$ is the confusion matrix, $N_{ij}$ is the number of vertices in common between community $C_i$ and $C_j$, $N_{.j}$ is the sum over column $j$ of $N$ and $N_{i.}$ is the sum over row $i$ of N. Note that the value of NMI ranges between $0$ and $1$, higher values represent more accurate results for an algorithm.

\subsection{Synthetic networks}
In this subsection, we perform a set of experiments on  Lancichinetti-Fortunato-Radicchi(LFR) benchmark networks\cite{Lancichinetti2008Benchmark}. Specifically, we set the average degree of LFR networks as 20, the maximum degree as 50, the exponent of the degree distribution as - 2.0, and the exponent of the community size distribution as - 1.0. With these parameters, we display four scenarios in the following: networks with 1000 nodes and community sizes changing from 10 to 50 nodes, which can be called 1000(S); networks with the same size but with community sizes varying from 20 to 100 nodes, which be called 1000(B); and the other two scenarios are subject to the same range of community size as the former two, but networks with 5000 nodes, which named as 5000(S) and 5000(B), respectively. The performance of the algorithms in experiment is quantified by NMI. In particular, it is worth pointed out that we generate 10 networks with the same parameters and take the average as final result in order to eliminate the randomness of benchmark networks. 
\begin{flushleft}
	\begin{figure}[!htb]
		\centering  
		\subfigure[LFR1000(S)]{\includegraphics[width=0.49\linewidth]{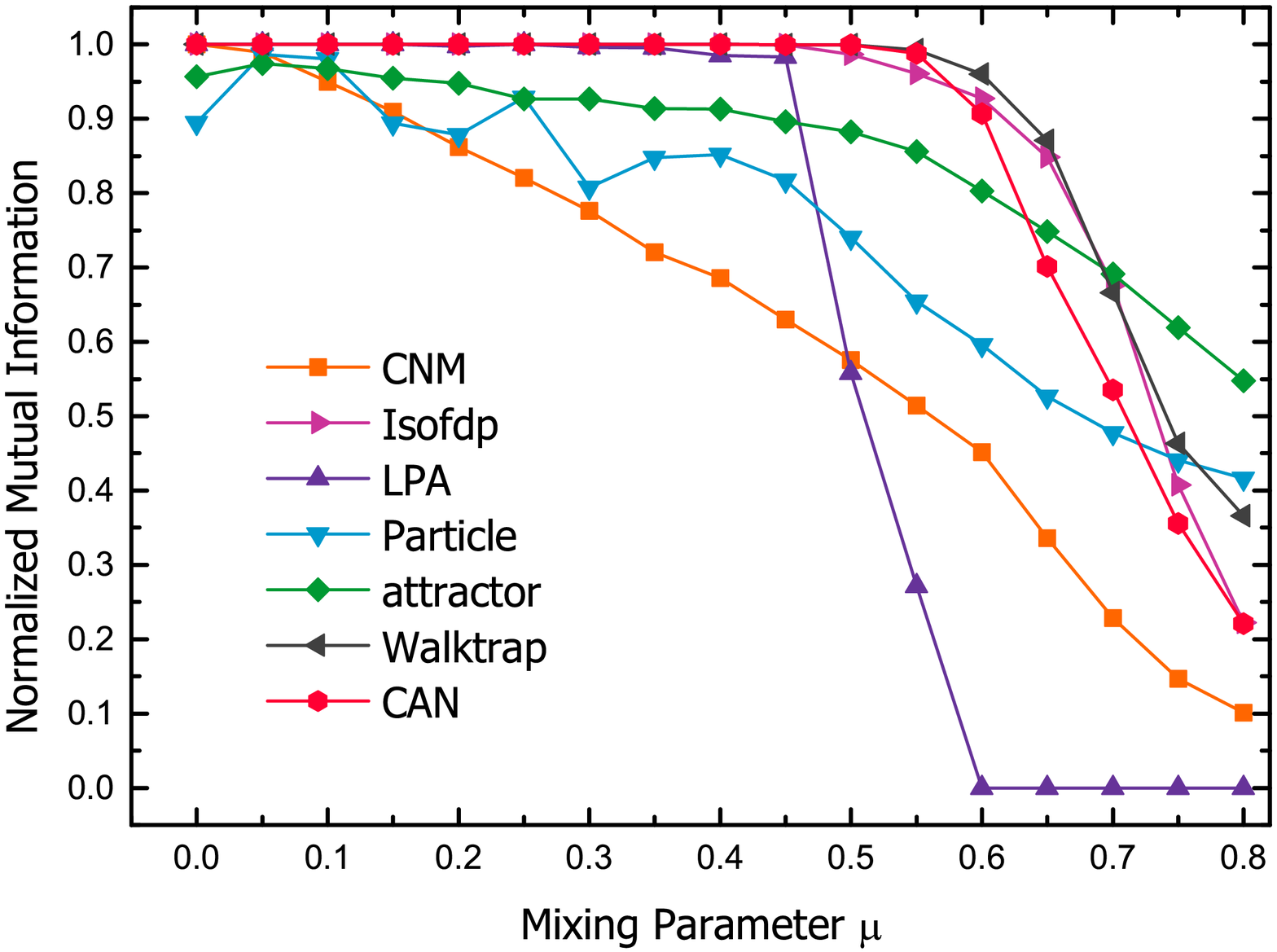}}
		\subfigure[LFR1000(B)]{\includegraphics[width=0.49\linewidth]{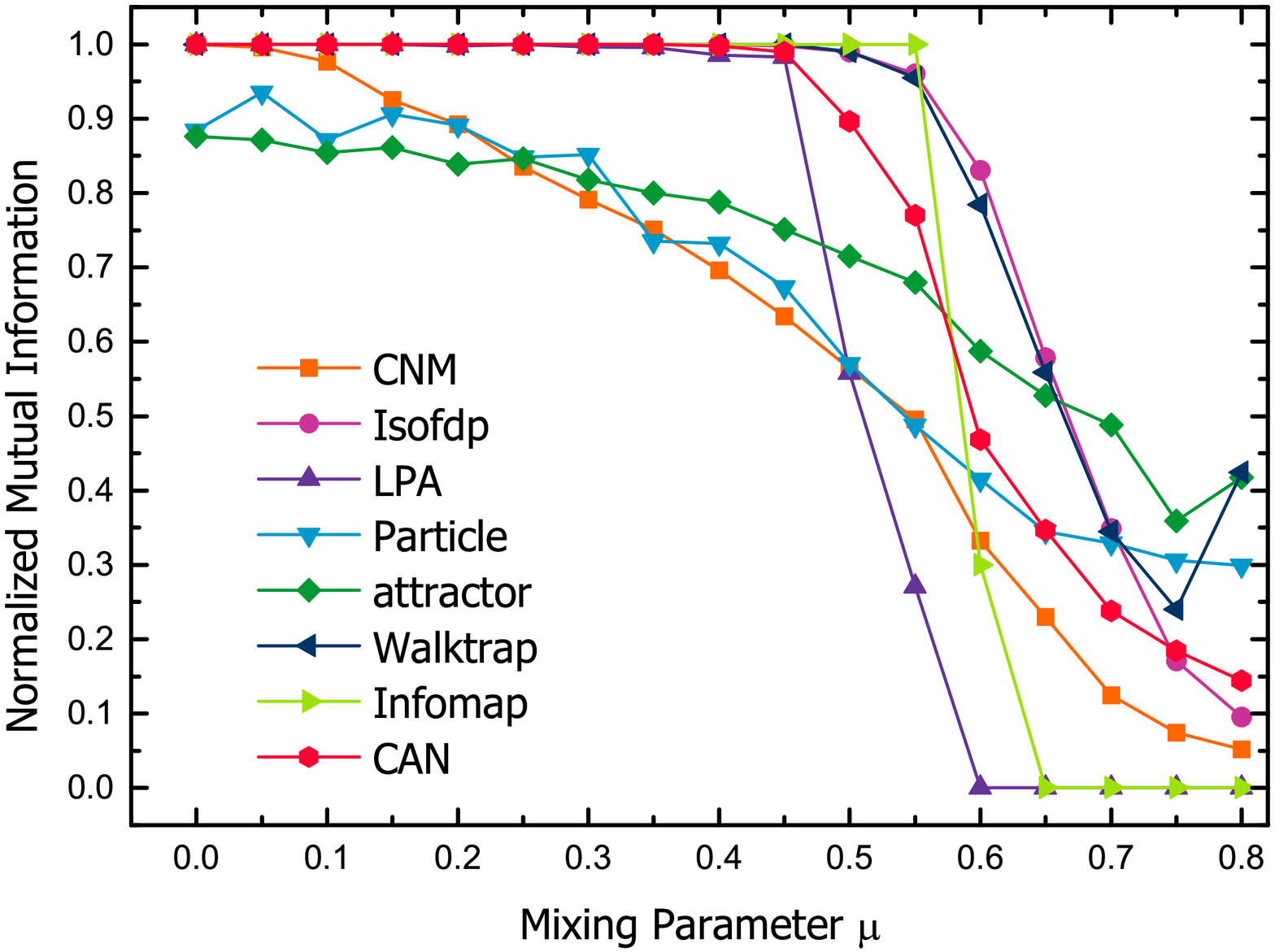}}
		\caption{Comparison of different community detection algorithms on LFR benchmark networks with N = 1,000. (a) Benchmark networks with communities of small size. (b) Benchmark networks with communities of big size. }
		\label{fig:LFR1000}
	\end{figure}
\end{flushleft}
As shown in Figure \ref{fig:LFR1000}, the proposed CAN algorithm gets $NMI = 1$ when $\mu <= 0.5$ on the networks with communities of small size, and gets $NMI = 1$ when $\mu <= 0.4$ on the networks with communities of big size, which means that the result matches with the natural network structures perfectly. Furthermore, in our experiments, we observe that CNM algorithm performs relatively bad compared to our algorithm on the small networks with two scenarios of community size. The LPA uncovers the ground truth community structures both on LFR1000(S) and LFR1000(B) when $\mu <= 0.45$, but its performance declines sharply as the $\mu$ increases. The main reason is that big communities are produced during label propagation when the boundary between communities is increasingly obscure. The Attractor and Particle descends slowly throughout the experiments on the small networks, but its effectiveness is always worse than the proposed algorithm when the mixing parameter $\mu <= 0.6$, and the result of Particle is subject to the randomness of benchmark networks, emerging slight fluctuation when $\mu < 0.4$. Moreover, Isofdp and  Walktrap achieve comparable performance to CAN algorithm, and even better than the proposed algorithm on LFR1000(S) when mixing parameter $\mu >= 0.6$ and on LFR1000(B) when $\mu >0.45$. Hence, we can conclude that our method works well and obtains better performance, compared to most algorithms.
\begin{flushleft}
	\begin{figure}[!htb]
		\centering  
		\subfigure[LFR5000(S)]{\includegraphics[width=0.49\linewidth]{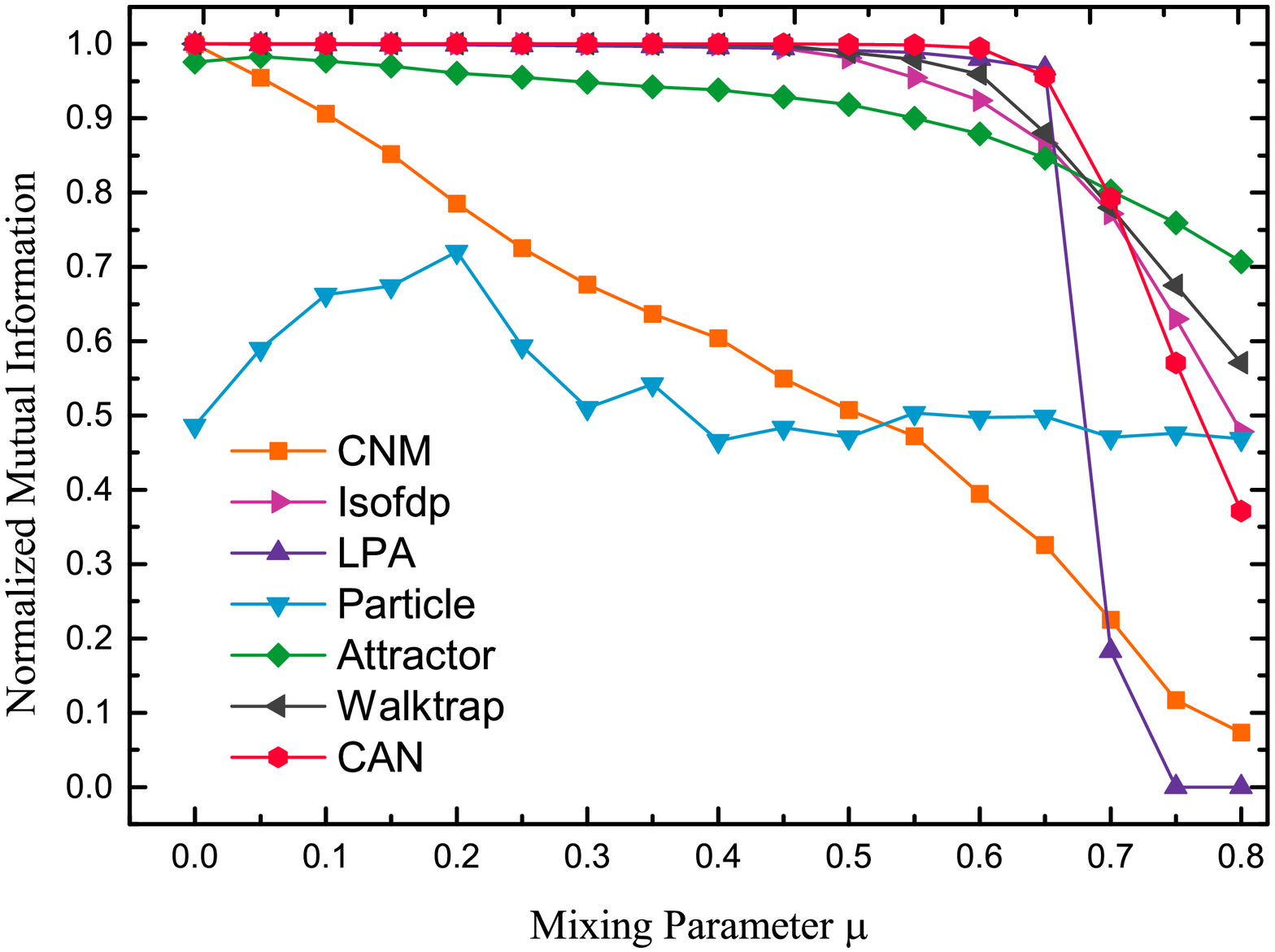}}
		\subfigure[LFR5000(B)]{\includegraphics[width=0.49\linewidth]{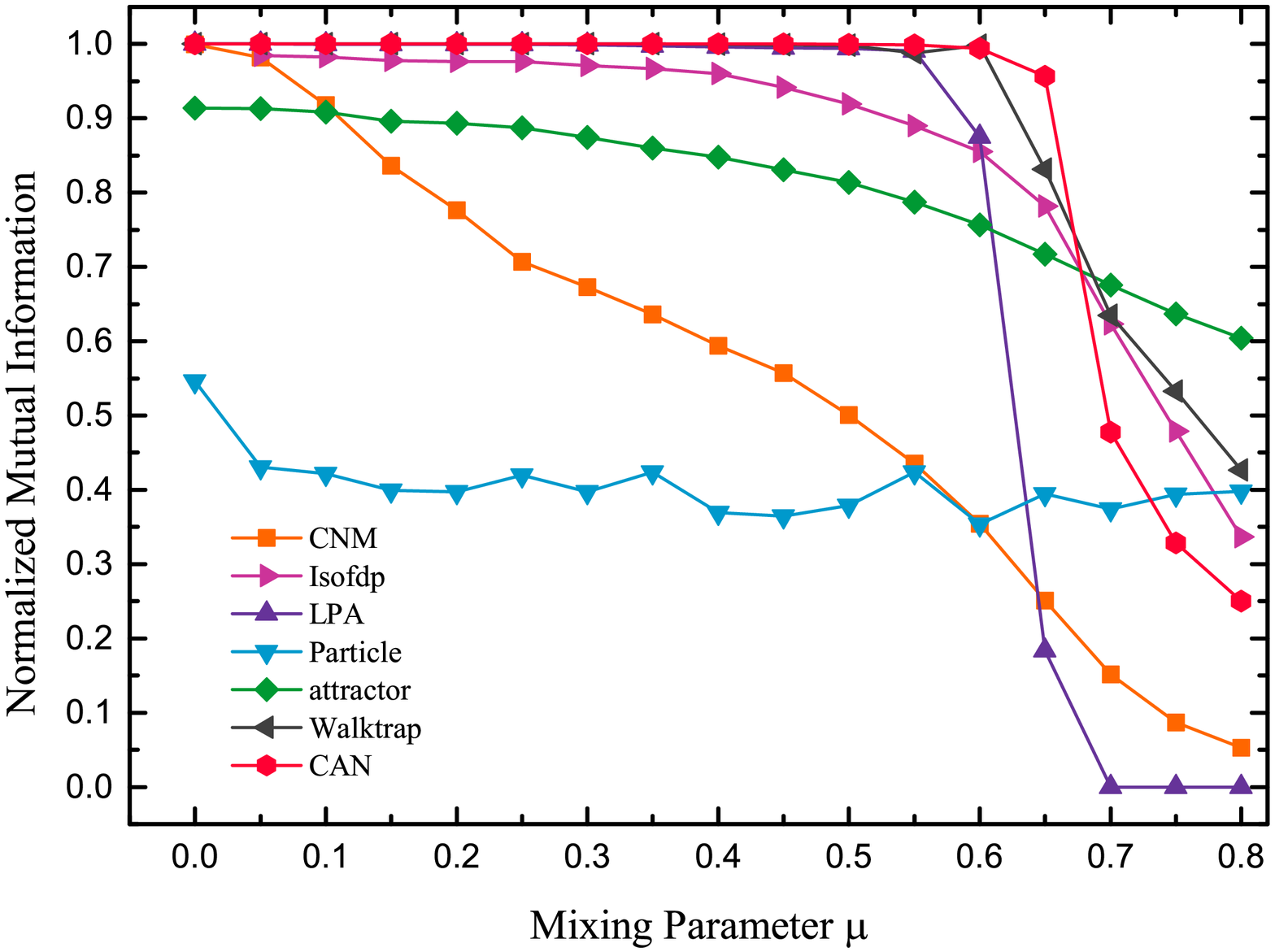}}
		\caption{Comparison of different community detection algorithms on LFR benchmark networks with N = 5,000. (a) Benchmark networks with communities of small size. (b) Benchmark networks with communities of big size. }
		\label{fig:LFR5000}
	\end{figure}
\end{flushleft}

As can be seen from Figure \ref{fig:LFR5000}(a), the proposed algorithm obtains optimal results when $\mu <= 0.7$ on the networks with communities of small size. Note that Isofdp, Attractor and Walktrap methods acquire relatively good results, and even better than CAN algorithm when $\mu > 0.7$. The CNM performs relatively bad compared to other algorithm. However, Particle algorithm is not better than CNM in general because of its sensitivity to randomness of LFR networks. The LPA disclose intrinsic community structure when $\mu <= 0.65$, but its performance declines sharply as the $\mu$ increases. On the other hand, the results of LFR5000(B) are depicted in Figure \ref{fig:LFR5000}(b), our algorithm is also superior when $\mu <= 0.65$. 

In summary, the proposed algorithm gets the superior performance on LFR benchmark networks, especially on large-scale networks.

\subsection{Real-world networks}\label{rwn}
In this subsection, we compare the performance of CAN with the compared algorithms on eight real-world networks -- Zechary's karate club network\cite{zachary1977information}, American college football network\cite{girvan2002community}, Scientists collaboration network\cite{girvan2002community}, Dolphin social network\cite{LusseauS186emergent}, Riskmap network\cite{Riskmap}, Political book network(Polbooks)\cite{newman2006modularity}, Les Miserables network\cite{knuth1993stanford} and PGP network\cite{PhysRevE.70.056122}. All these networks are commonly used in community detection. The basic information of these networks are shown in Table \ref{tab:networks}.

\begin{table}[htbp] 
	\centering
	\caption{The basic information of the real-world networks}
	\label{tab:networks}
	\begin{tabular}{lcccc}
		\hline
		\textbf{Network} & \textbf{vertices} & \textbf{edges} & $\langle\textbf{k}\rangle$ & \textbf{communities}\\
		\hline
		Karate  & 34       &  87   & 4.59 & 2\\
		Football& 115	   &  613  & 10.66 & 11\\
		SantaFe & 118	   &  197  & 3.34  & 6\\
		Dolphins& 62       & 159   & 5.13  & 4\\
		Rsikmap & 42       & 83    & 3.92  & 6\\
		Polbooks& 105      & 441   &  8.40 & 3\\
		Lesmis  & 77       & 253   & 6.57 & 7\\
		PGP & 10680        & 24316 &  4.55 & 713\\
		\hline
	\end{tabular}
\end{table}	

\begin{figure}[!htb]
	\makebox[\textwidth][c]{
		\includegraphics[width=0.65\textwidth]{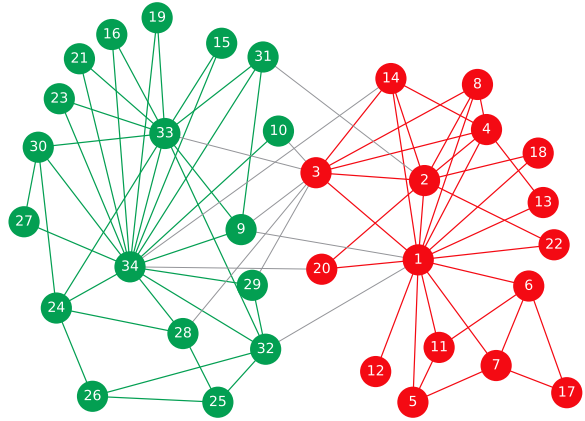} }
	\caption{The community structure of the Zachary's Karate Club network detected by CAN algorithm.}\label{fig:karate}
\end{figure}

The Zechary's karate club network contains 34 nodes, and partitions into two smaller clubs ultimately after a dispute between the instructor (Vertex 34) and the administrator(Vertex 1). As shown in Figure \ref{fig:karate}, there are two communities obtained by our proposed algorithm for $\beta \in [0.43,0.62]$, which coincide with ground truth community structure. The comparison results of the two metrics are listed in Table \ref{tab:compare}

\begin{flushleft}
	\begin{figure}[!htb]
		\centering  
		\subfigure[]{\includegraphics[width=0.49\linewidth]{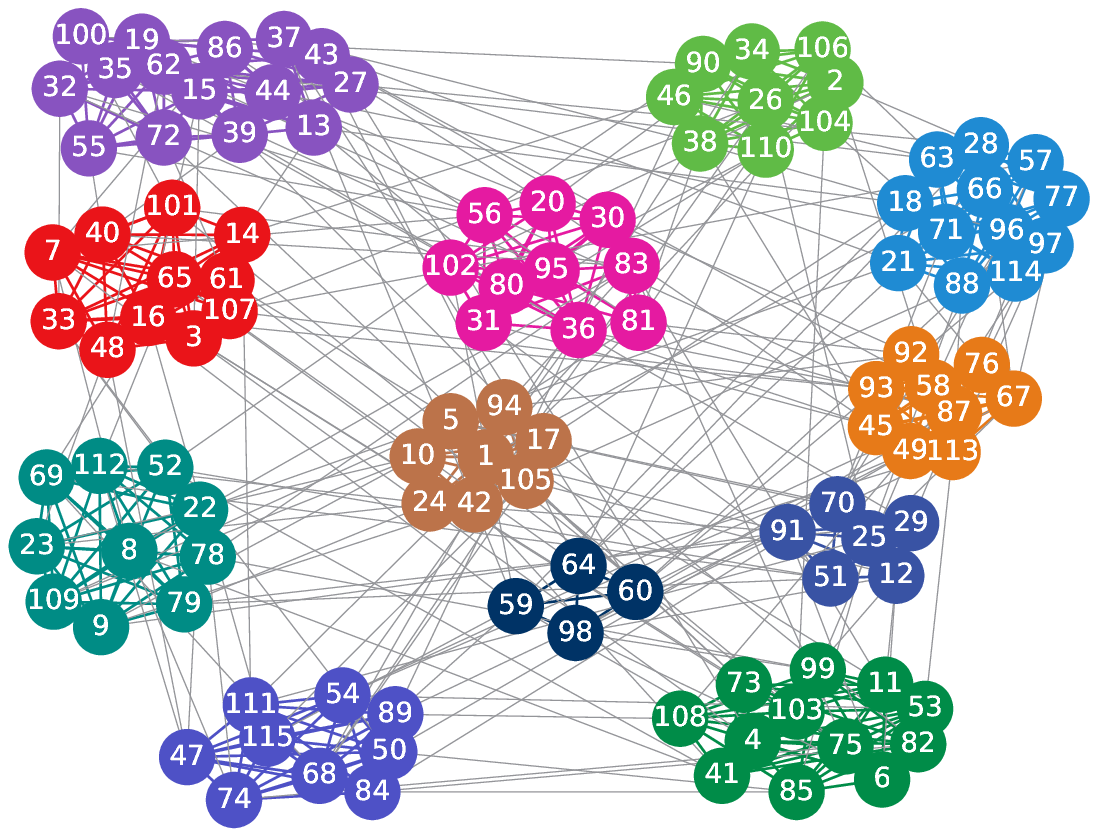}}
		\subfigure[]{\includegraphics[width=0.49\linewidth]{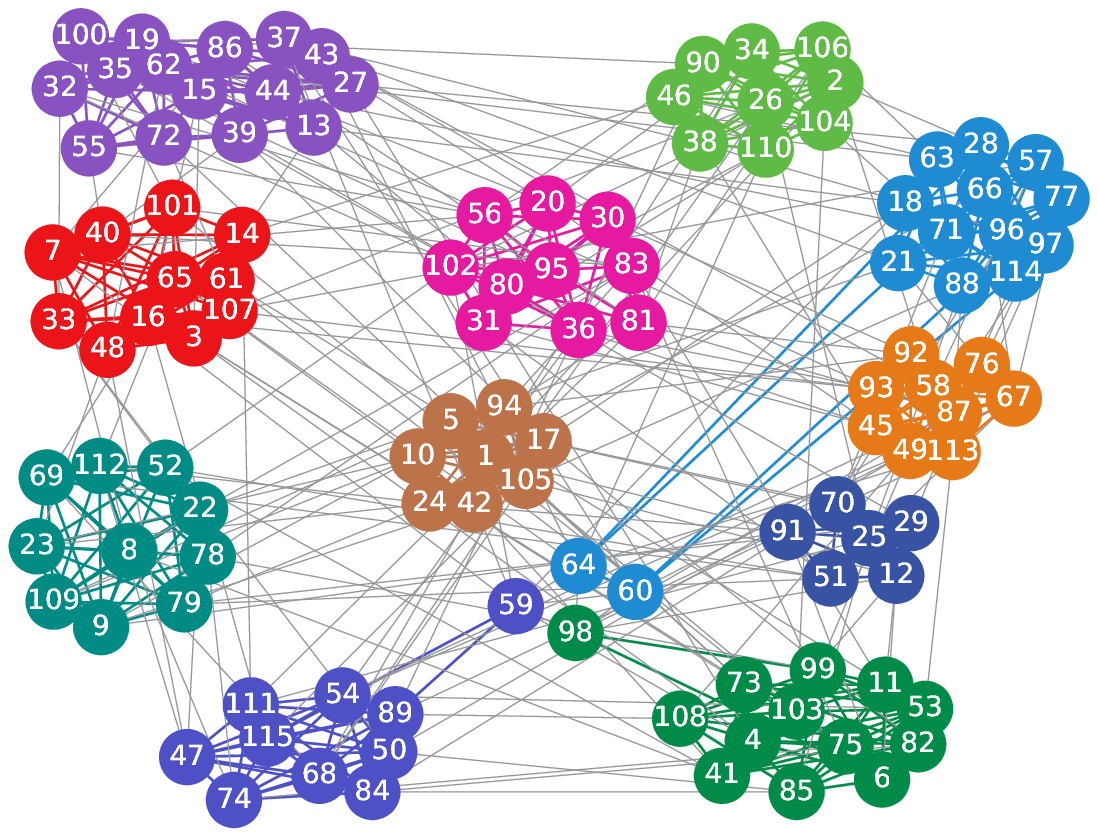}}
		\caption{The result of CAN algorithm on American College Football network.(a)is the ground truth;(b) is the obtained network partition by CAN.}
		\label{fig:football}
	\end{figure}
\end{flushleft}

The American college football network represents the schedule of games between American college football teams during a regular season. In the network nodes represent the 115 teams that are divided into 12 groups, and the edges denote 616 games. The groud truth community structure is shown in Fig.\ref{fig:football}. As can be seen from Fig. \ref{fig:football}(b), our Algorithm uncovers 11 communities for $\beta \in [0.67,0.81]$. The conference \{59, 98, 60, 64\} has been divided into three partitions. This is the reason that the team in this conference does not satisfy the assumption that  two teams in the same conference would have more  games than two in different conference. For example, the neighbors of node 59 are $\{60, 18, 37, 115, 98, 89, 64, 4, 7, 102\}$, it means that team 59 has only three matches in its conference and seven games with the other conferences. So the result obtained by our method is consistence with the real back ground. The comparison results of the two metrics are enumerated in Table \ref{tab:compare}.

\begin{flushleft}
	\begin{figure}[!htb]
		\centering  
		\subfigure{\label{fig:gtsantafe}\includegraphics[width=0.49\linewidth]{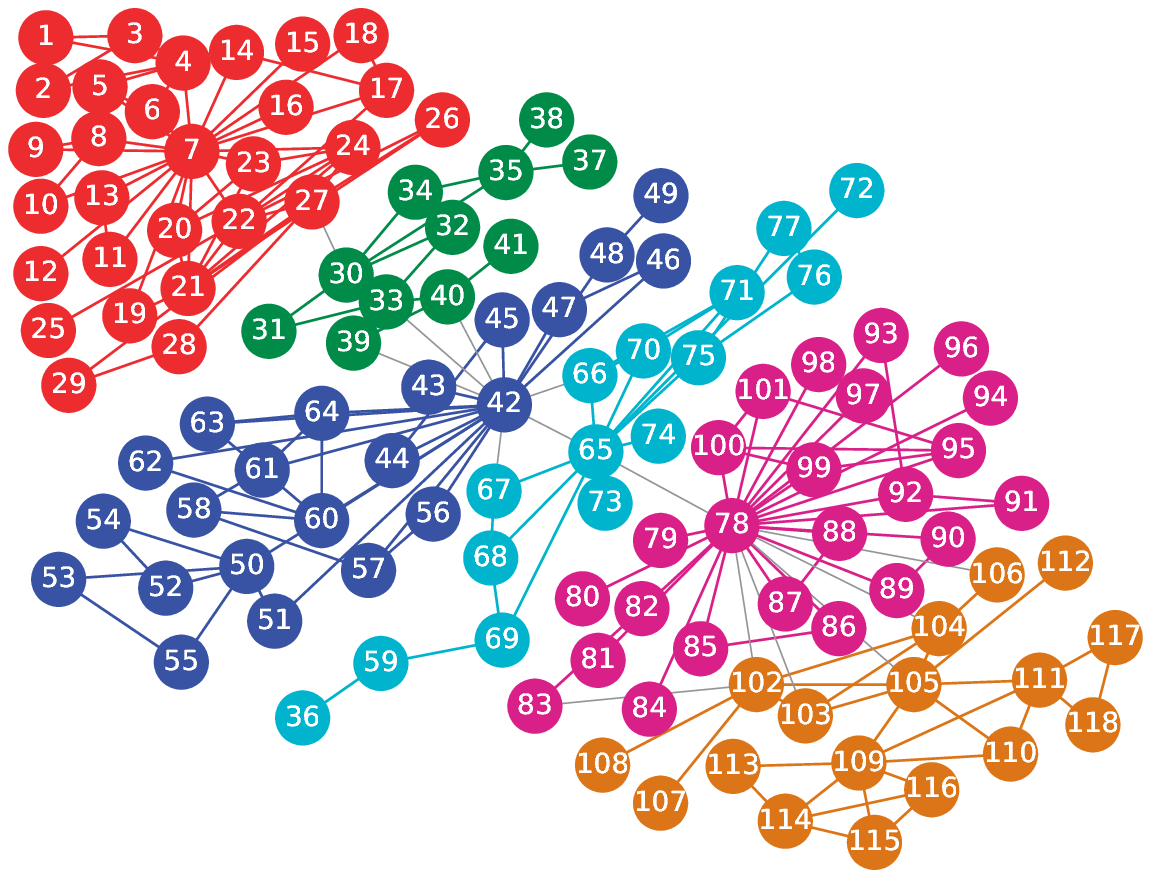}}
		\subfigure{\label{fig:santafe}\includegraphics[width=0.49\linewidth]{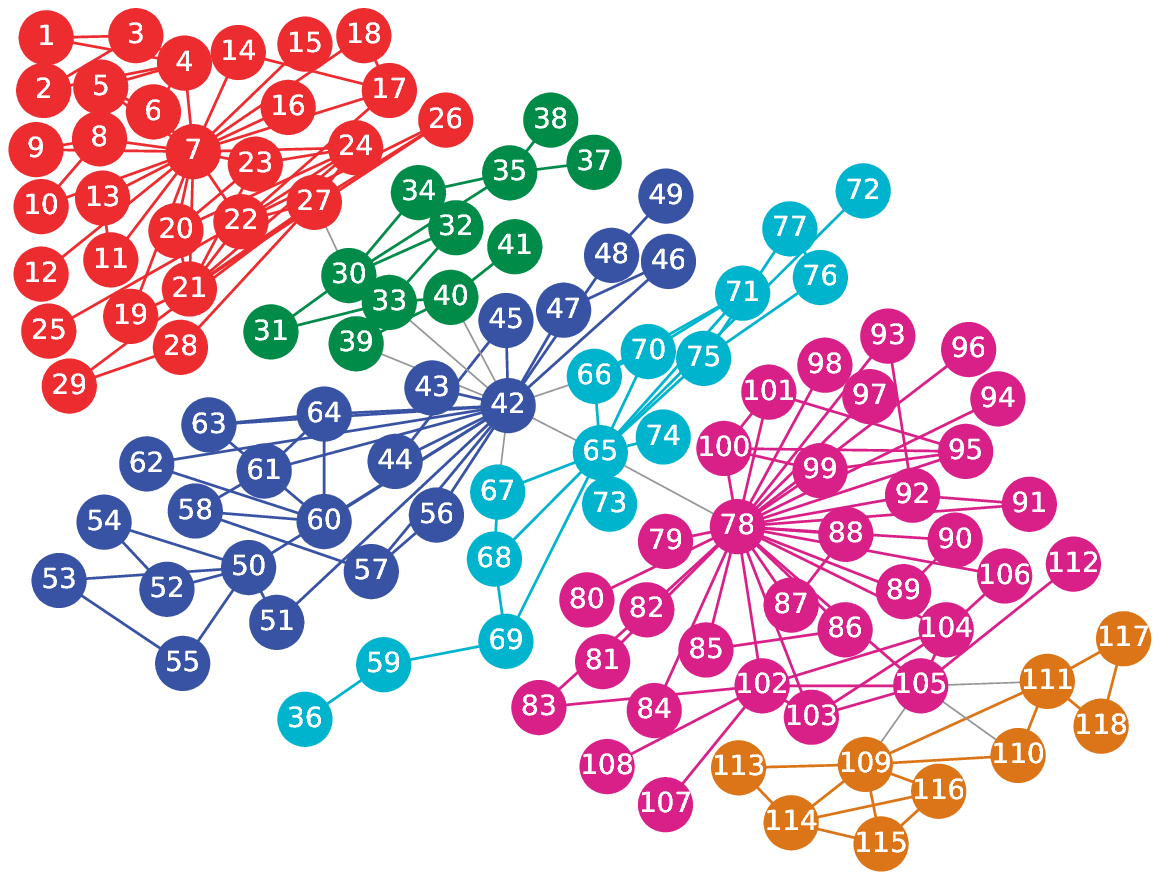}}
		\caption{The result of CAN algorithm on SantaFe network. (a) is ground truth. (b) is identified by the proposed algorithm.}
	\end{figure}
\end{flushleft}

The scientists collaboration network is a network depicting the co-author relationship between scientists at the Santa Fe Institute. This network consists of  118 nodes and 197 edges. An edge represents two nodes(scientists) coauthored one or more articles during the same period. As shown in Fig.\ref{fig:gtsantafe}, the ground truth community structure contains 6 communities. Note that the proposed algorithm also split it into 6 communities for $\beta \in [0.72, 0.74]$, which is illustrated in Figure \ref{fig:santafe}. Obviously, the node set \{102, 103, 104, 105, 106, 107, 108, 112\} detected by CAN are assigned to incorrect communities compared with the ground truth community structure. Some of them, including \{102, 103, 104, 105, 106\}, are located at community boundary between Mathematical Ecology(Pink color) and Agent-based Models(Mandarin color). Among them, there exists a node with critical influence who takes a part of the center of gravity, that is node 78. In the process of finding a pair of nodes with strong correlation, these boundary nodes are strong correlation with node 78. Hence, these marginal nodes tend to be misclassified into community of Mathematical Ecology, and the misclassification of the other nodes {107, 108, 112} are inevitable. However, the communities identified by CAN algorithm is acceptable and considerably superior compared with the other community detection algorithm in Table \ref{tab:compare}.     

\begin{flushleft}
	\begin{figure}[!htb]
		\centering  
		\subfigure[]{\label{fig:gtdolphins}\includegraphics[width=0.49\linewidth]{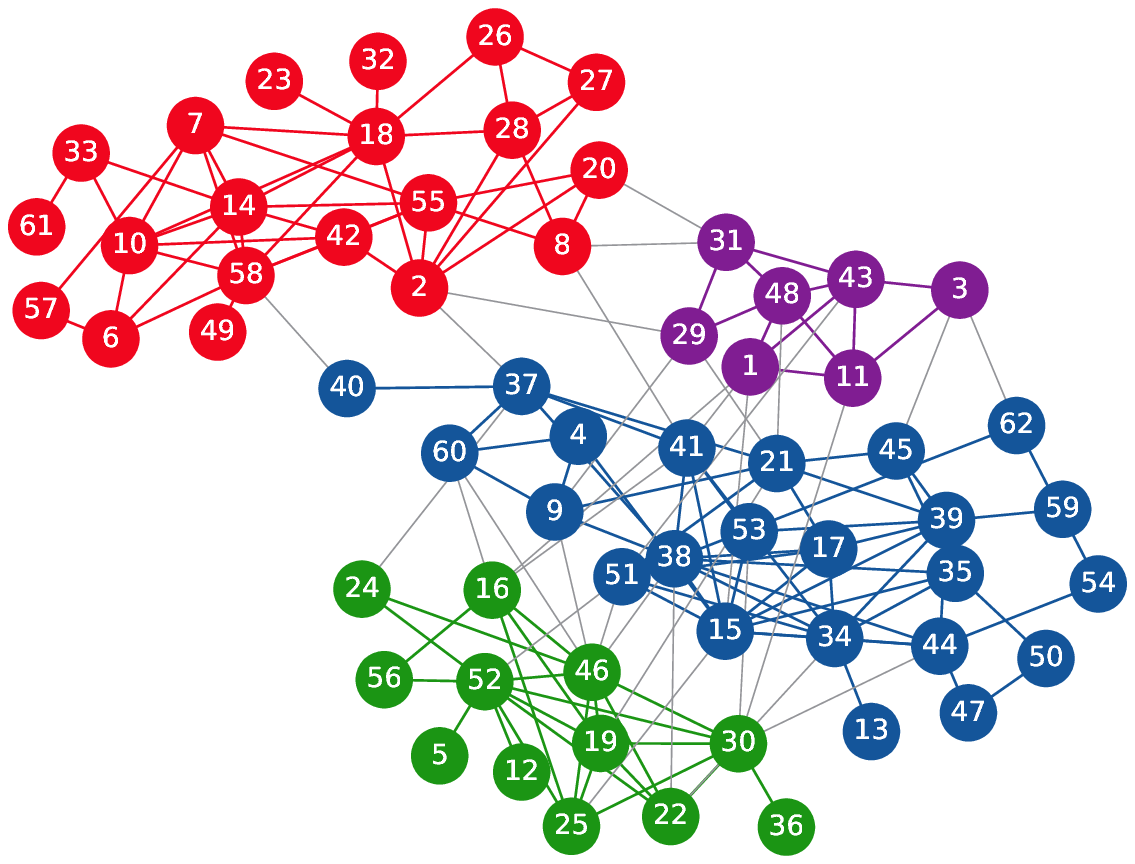}}
		\subfigure[]{\label{fig:dolphins}\includegraphics[width=0.49\linewidth]{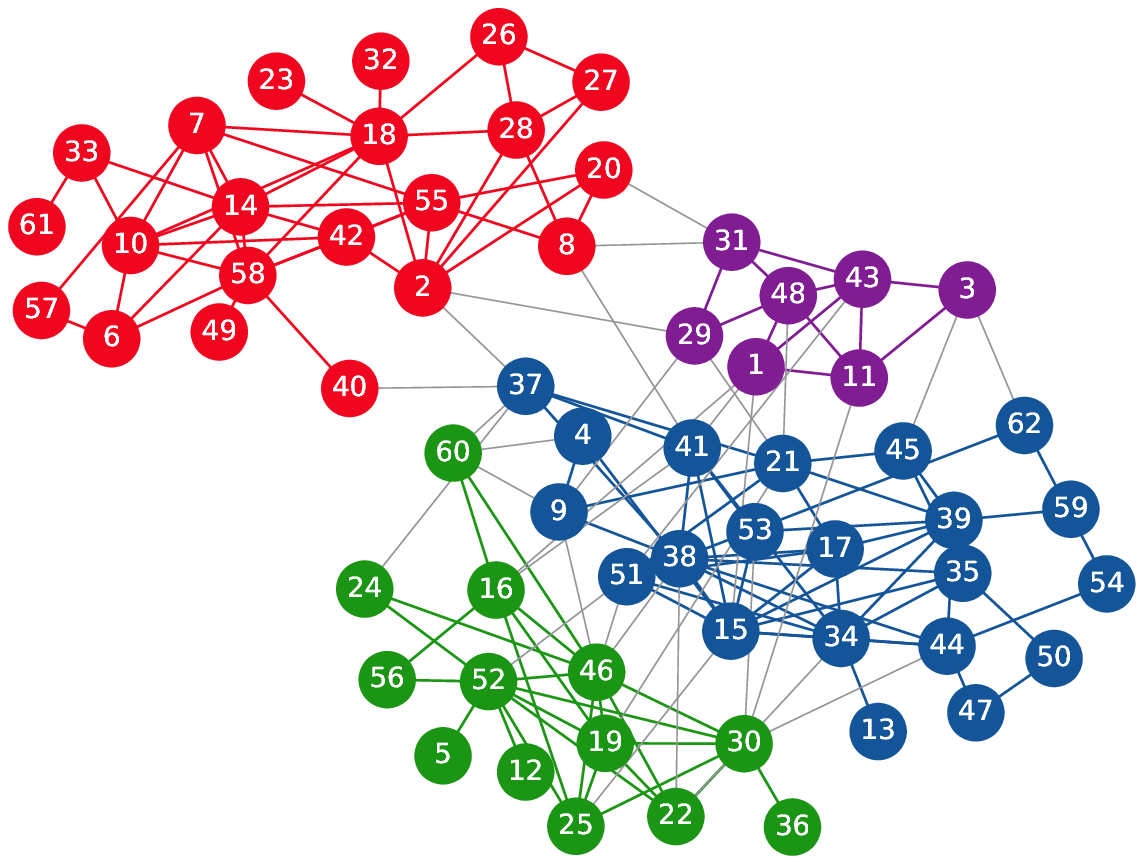}}
		\caption{The result of CAN algorithm on dolphin social network. (a) is ground truth. (b) is identified by the proposed algorithm.}
	\end{figure}
\end{flushleft}

The dolphin social network was constructed from observations recording frequent associations between a community of 62 bottlenose dolphins over a period of 7 years from 1994 to 2001. In this network, nodes represent dolphins and edges between them donates that they often interact with each other. In the previous work, it is generally divided into two groups or four sub-groups in the light of sex and age of dolphins. The ground truth community structure of Dolphins social network is illustrated in Fig.\ref{fig:gtdolphins}. As shown in Figure \ref{fig:dolphins}, our method reveals four communities for $\beta = 0.68$ that are remarkably close to the ground truth, while nodes 'sn89'(40) and 'zap'(60) have been misclassified into adjacent communities. Note that nodes 'sn89' and 'zap' is located at the community boundary. Thus, it is easy to understand why it is assigned into wrong community. The comparison results of the two evaluation metrics are listed in Table \ref{tab:compare}. 

\begin{flushleft}
	\begin{figure}[!htb]
		\centering  
		\subfigure[]{\label{fig:gtriskmap}\includegraphics[width=0.49\linewidth]{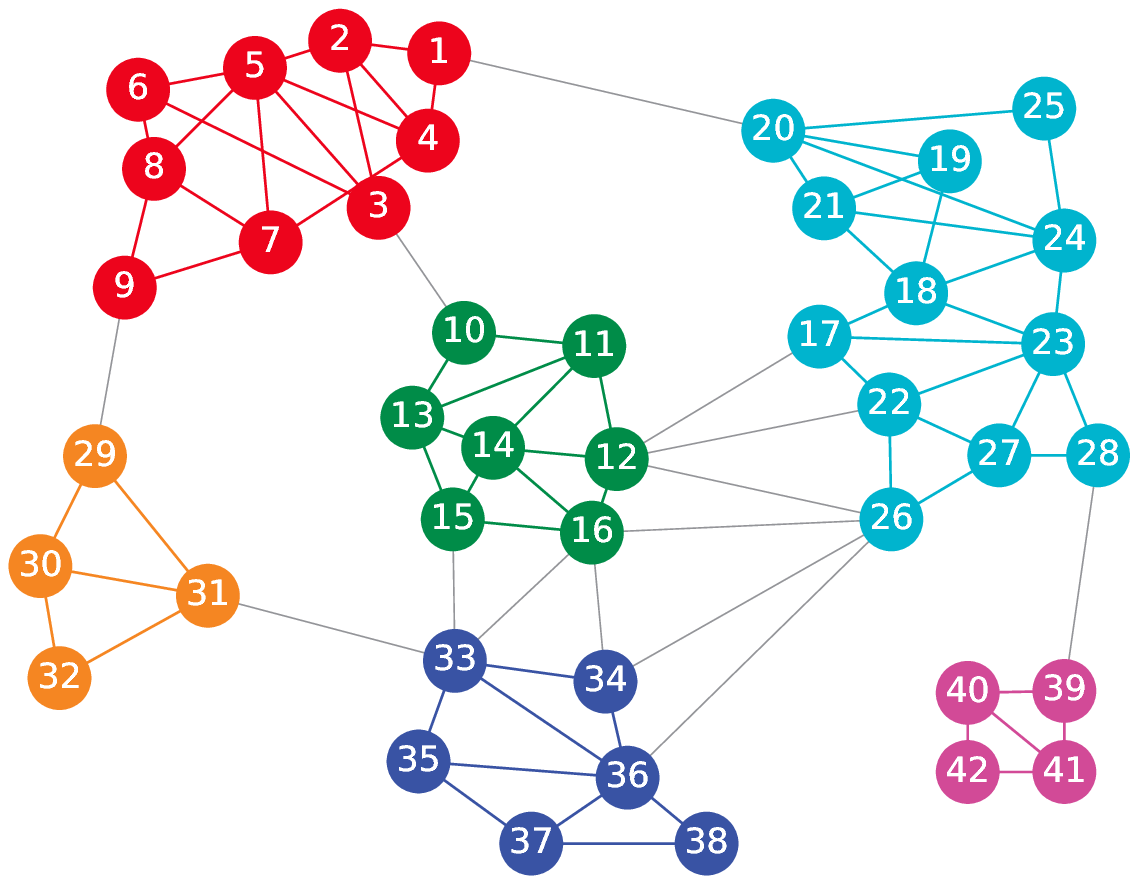}}
		\subfigure[]{\label{fig:riskmap}\includegraphics[width=0.49\linewidth]{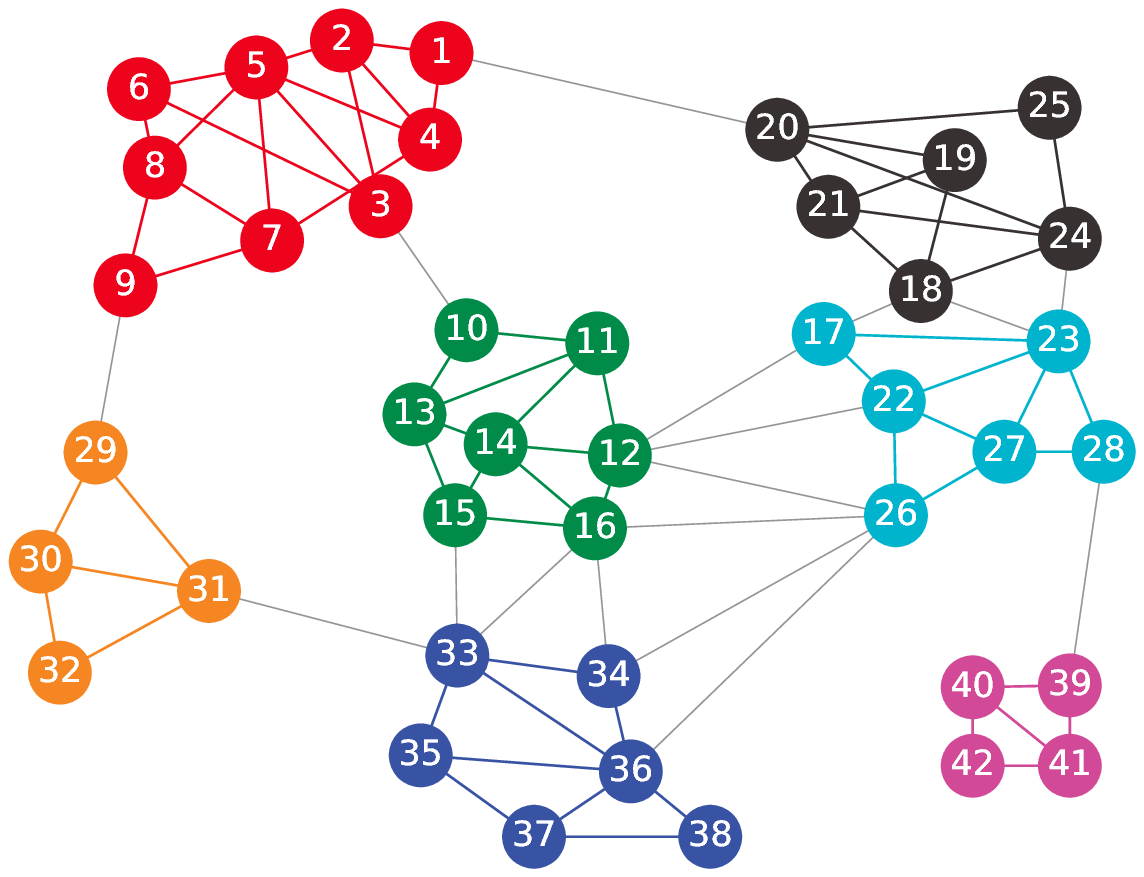}}
		\caption{The result of  CAN algorithm on Riskmap network. (a) is the ground truth community structure; (b) is detected by the CAN algorithm.}
	\end{figure}
\end{flushleft}

Riskmap network is created according to a map of popular board game Risk, which was invented in 1957 by Albert Lamorisse. It contains 42 nodes that represent territories and 83 edges donate pairs of territories are geographical adjacency. To avoid any political sensitivity, each of the nodes are treated as a continuous number rather than the name of the country or territory, the ground truth of which consists of 6 communities shown in Fig. \ref{fig:gtriskmap}. The community detected by our proposed algorithm includes 7 communities for $\beta \in [0.83,0.84]$ depicted in Fig. \ref{fig:riskmap}. Obviously, the community \{17, 18, 19, 20, 21, 22, 23, 24, 25, 26, 27, 28\} has been divided into two sub-communities, \{18, 19, 20, 21, 24, 25\} and \{17, 22, 23, 26, 27, 28\}. Firstly, we observe that the edges between two sub-communities are sparse than intra-communities. Moreover, we find that two sub-communities, in and of itself, are equivalence classes. Thus, it is reasonable that the Riskmap network partitions into 7 communities. The comparison results of NMI and Modularity are listed in Table \ref{tab:compare}.

\begin{flushleft}
	\begin{figure}[!htb]
		\centering  
		\subfigure[]{\label{fig:gtpolbooks}\includegraphics[width=0.49\linewidth]{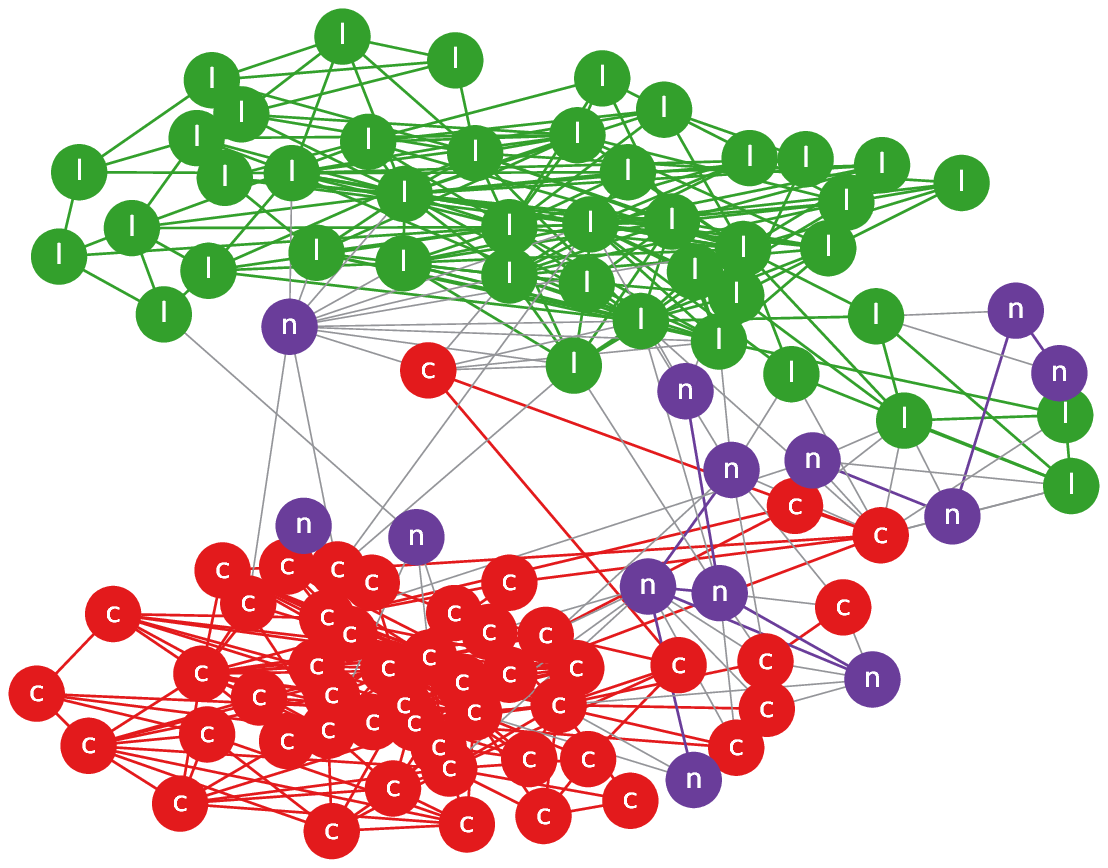}}
		\subfigure[]{\label{fig:polbooks}\includegraphics[width=0.49\linewidth]{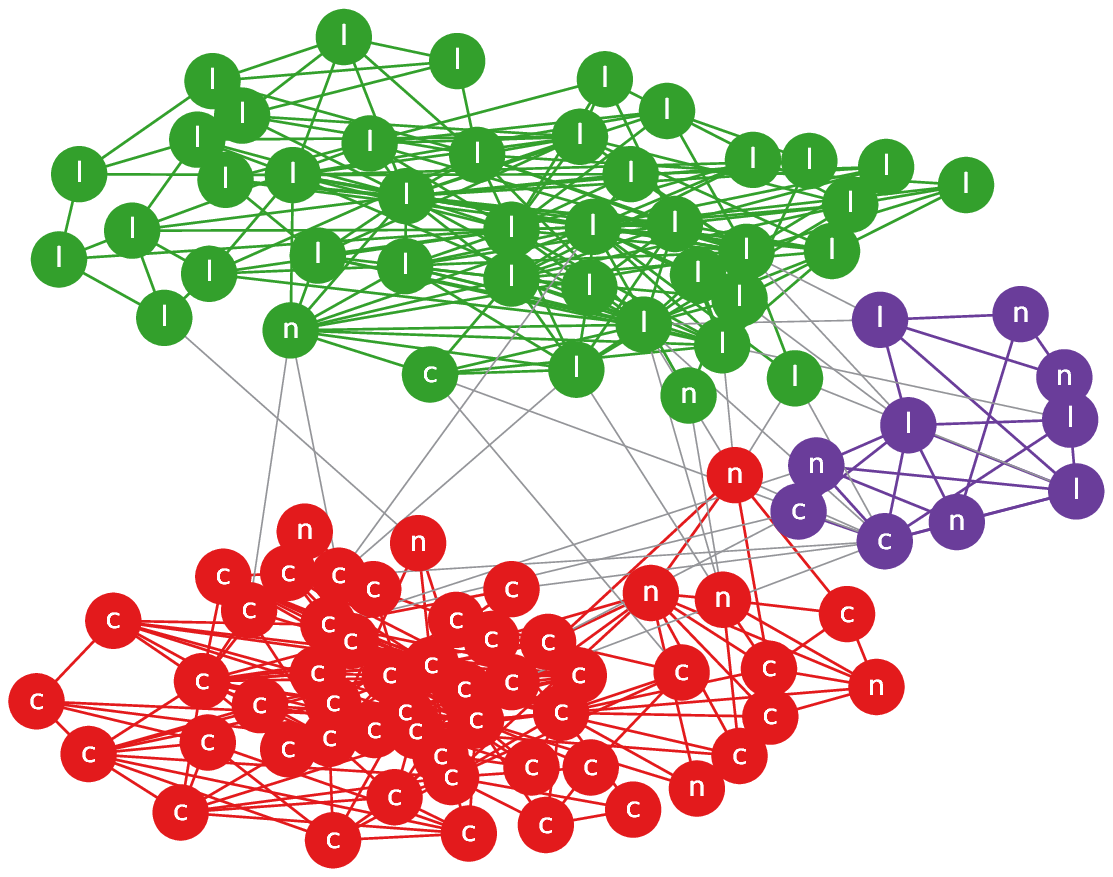}}
		\caption{The result of  CAN algorithm on political book network. (a) is the ground truth community structure; (b) is detected by the CAN algorithm.}
	\end{figure}
\end{flushleft}

Political book network contains 105 nodes that donate books about US politics sold by the online bookseller Amazon.com. Edges represent frequent co-purchasing of books by the same customers. Each book is labeled with "liberal", "neutral" and "conservation", respectively, which are represented as 'l', 'n' and 'c' in our experiment. As can be seen from Figure \ref{fig:gtpolbooks}, the ground truth contains three communities. As show in Fig.\ref{fig:polbooks}, CAN algorithm also partitions these books into three categories, where two communities well signify the corresponding liberal and conservative books, separately. Compared with the ground truth, the communities detected by our method are closer to the definition of community, while the NMI measure of the proposed algorithm is smaller than Particle Algorithm shown in Table \ref{tab:compare}. 

The Les Miserables network describes the interactions between major figures in the Victor Hugo's novel "Les Miserables", as compiled by Knuth. Nodes represent characters as indicated by the labels and edges connect any pair of characters that appear in the same chapter of the book. As shown in Table \ref{tab:compare}, CAN obtains the best community quality comparing to other well-known algorithm when using the internal criteria of Modularity. 

The PGP network is a trust networks on base of an encryption program, where vertices are certificates and an edge donates authorization from the owner of a certificate to that of another. As shown in Table \ref{tab:compare}, the proposed algorithm gets the maximum modularity with the exception of the result of CNM algorithm. It means that our method is also available for large-scale networks.

\renewcommand\arraystretch{1.7}
\begin{table}\addtolength\tabcolsep{-0.23em}
\centering
\scriptsize
\caption{Comparisons of the 2 metrics on real-world networks: A rank(number in parentheses) is attached to the value of each metric for each network, and the value of score means the average rank of two metrics. R represents the final rank of each algorithm.}
\label{tab:compare}
\begin{tabular}{cccccc||cccccc}\hline
	Dataset       & Alg. & Q & NMI & Score & R  & Dataset & Alg. & Q & NMI & Score & R \\
	\hline
	\multirow{7}{*}{Karate} & CNM & 0.381(1) & 0.693(3) & 2 & 2 & \multirow{7}{*}{Football} & CNM & 0.550(6) & 0.751(6) & 6 & 5\\
	& Isofdp & 0.372(2) & 1.000(1) & 1.5 & \textbf{1} & & Isofdp & 0.600(3) & 0.982(2) & 2.5 & 2 \\
	& LPA & 0.354(3) & 0.622(4) & 3.5 & 3 & &  LPA & 0.589(4) & 0.945(5) & 4.5 & 4\\
	& Walktrap & 0.353(4) & 0.504(5) & 4.5 & 4 & & Walktrap & 0.603(1) & 0.954(4) & 4 & 3\\
	& Particle & 0.064(5) & 0.161(6) & 5.5 & 5 & & Particle & 0.585(5) & 0.954(4) & 4.5 & 4\\
	& Attractor & 0.372(2) & 0.924(2) & 2 & 2 & & Attractor & 0.601(2) & 0.989(1) & 1.5 & \textbf{1} \\
	& CAN & 0.372(2) & 1.000(1) & 1.5 & \textbf{1} & & CAN & 0.601(2) & 0.970(3) & 2.5 & 2 \\ \hline
	\multirow{7}{*}{Santafe} & CNM & 0.750(1) & 0.867(2) & 1.5 & \textbf{1} & \multirow{7}{*}{Dolphins} & CNM & 0.492(2) & 0.733(2) & 2 & 2\\
	& Isofdp & 0.668(5) & 0.825(4) & 4.5 & 4 & & Isofdp & 0.479(4) & 0.683(5) & 4.5 & 4 \\
	& LPA & 0.638(6) & 0.741(6) & 6 & 4 & &  LPA & 0.457(5) & 0.711(3) & 4 & 3\\
	& Walktrap & 0.733(3) & 0.818(5) & 4 & 3 & & Walktrap & 0.489(3) & 0.632(6) & 4.5 & 4\\
	& Particle & 0.587(7) & 0.575(7) & 7 & 5 & & Particle & 0.408(7) & 0.512(7) & 7 & 6\\
	& Attractor & 0.694(4) & 0.836(3) & 3.5 & 2 & & Attractor & 0.443(6) & 0.699(4) & 5 & 5 \\
	& CAN & 0.738(2) & 0.927(1) & 1.5 & \textbf{1} & & CAN & 0.519(1) & 0.903(1) & 1 & \textbf{1} \\ \hline
	\multirow{7}{*}{Riskmap} & CNM & 0.625(2) & 0.894(3) & 2.5 & 2 & \multirow{7}{*}{Polbooks} & CNM & 0.502(2) & 0.531(5) & 3.5 & 3\\
	& Isofdp & 0.519(6) & 0.714(7) & 6.5 & 5 & & Isofdp & 0.483(4) & 0.443(7) & 5.5 & 5 \\
	& LPA & 0.597(5) & 0.830(5) & 5 & 4 & &  LPA & 0.488(3) & 0.529(6) & 4.5 & 4\\
	& Walktrap & 0.624(3) & 0.848(4) & 3.5 & 3 & & Walktrap & 0.507(1) & 0.543(4) & 2.5 & \textbf{1}\\
	& Particle & 0.621(4) & 1.000(1) & 2.5 & 2 & & Particle & 0.415(5) & 1.000(1) & 3 & 2\\
	& Attractor & 0.462(7) & 0.777(6) & 6.5 & 5 & & Attractor & 0.495(3) & 0.567(3) & 3 & 2 \\
	& CAN & 0.634(1) & 0.945(2) & 1.5 & \textbf{1} & & CAN & 0.495(3) & 0.586(2) & 2.5 & \textbf{1} \\ \hline
	\multirow{7}{*}{Lesmis} & CNM & 0.499(5) &  & 5 & 5 & \multirow{7}{*}{PGP} & CNM & 0.850(1) & & 1 & \textbf{1}\\
	& Isofdp & 0.510(4) &  & 4 & 4 & & Isofdp & 0.726(5) &  & 5 & 5 \\
	& LPA & 0.513(3) & & 3 & 3 & &  LPA & 0.769(4) & & 4 & 4\\
	& Walktrap & 0.519(2) &  & 2 & 2 & & Walktrap & 0.789(3) &  & 3 & 3\\
	& Particle & 0.382(7) &  & 7 & 7 & & Particle & 0.100(7) &  & 7 & 7\\
	& Attractor & 0.407(6) &  & 6 & 6 & & Attractor & 0.673(6) & & 6 & 6 \\
	& CAN & 0.524(1) &  & 1 & \textbf{1} & & CAN & 0.803(2) & & 2 & 2 \\
	\hline
\end{tabular}
\end{table}		

In summary, as shown in Table \ref{tab:compare}, the Rank of CAN algorithm gets first on six networks and takes second place on the other two networks. Obviously, the proposed algorithm shows superior performance, compared with the other algorithms. 
 
\subsection{Discussion}\label{discussion}
In our experiment, we observe that there are two parameter to be set. In fact, however, the default setting of $\lambda$ is 4, and we only need to adjust the threshold of correlation coefficient $\beta$ in all experiments.
\begin{figure}[!htb]
	\makebox[\textwidth][c]{
		\includegraphics[width=0.9\textwidth]{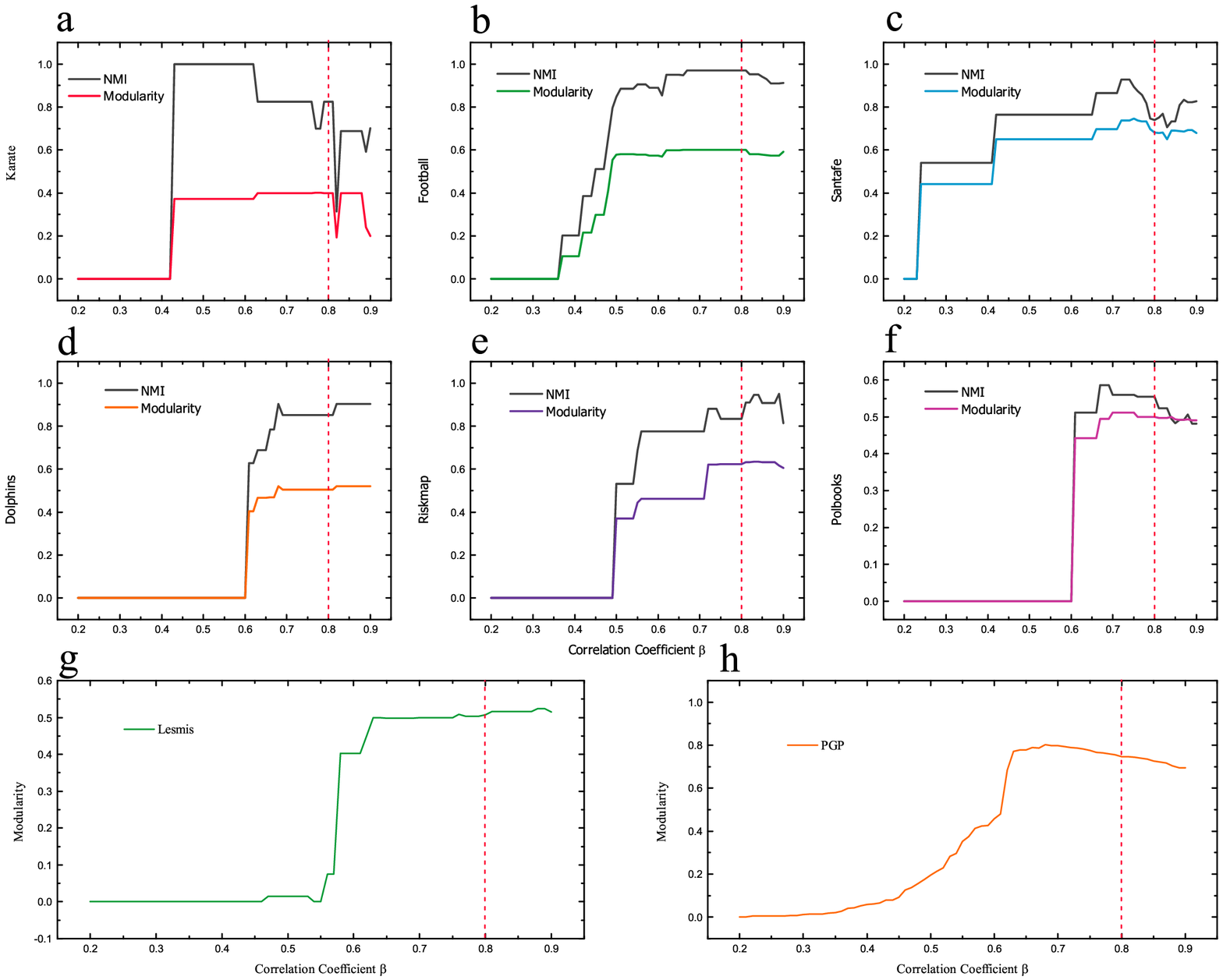} }
	\caption{The setting of parameter $\beta$ in experiments. \textbf{a-f} represents the relationship between correlation coefficient $\beta$ and two measurements in real-world networks with ground truth community structure. \textbf{g} shows the variation of modularity as the $\beta$ increases on Les Miserables network. \textbf{h} donates the relationships between Modularity and Correlation coefficient $\beta$ on PGP network.}\label{fig:ParameterSet}
\end{figure}

 As can be seen from Fig.\ref{fig:ParameterSet}, \textbf{a-f} reveal that the NMI have the same changing tendency with the Modularity in real-world networks with ground truth community structure as the correlation coefficient $\beta$ varying from 0.2 to 0.9, and we find that the proposed algorithm can obtain an acceptable result around the $\beta=0.8$. \textbf{g} shows the relationship between the correlation coefficient $\beta$ and Modularity on Les Miserables network. Obviously, Lesmis network also obtains a superior result when $\beta=0.8$. Moreover, \textbf{h} represents that CAN algorithm get an acceptable Modularity around $\beta=0.8$. Above all, the parameter $\beta$ can be selected from around of $0.8$.

\begin{figure}[!htb]
	\makebox[\textwidth][c]{
		\includegraphics[width=0.9\textwidth]{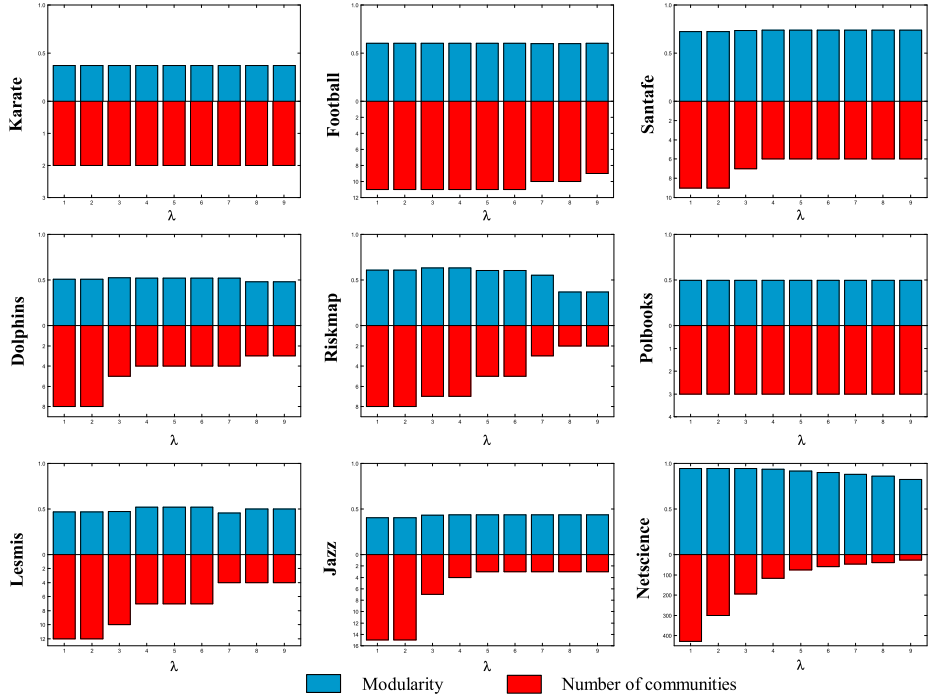} }
	\caption{The impacts of $\lambda$ on real-world networks.The X-axis is the parameter $\lambda$, and the Y-axis represents the modularity and number of communities.}\label{fig:lambda}
\end{figure}

We also analyze why the  parameter $\lambda$ is set as 4 in Fig. \ref{fig:lambda}. The number of communities drops gradually, but the change of modularity value is not obvious, which donates that the communities unsatisfied the criterion of community can be merged into the large-scale communities reasonably. The default setting for the parameter $\lambda$ is 4. The main reason is that the proposed algorithm can obtain a similar number of communities with the ground truth and get a reasonable value of Modularity and NMI.
\section{Conclusion}\label{conclusion}
In the paper, a novel CAN algorithm is proposed to reveal community structure using the correlation analysis of nodes. The main characteristic of the proposed algorithm is that it is based on the simple idea that nodes in the same community are more correlated than between communities. With regard to the correlation of nodes, we first construct the link pattern of each node. Then we find all pairs of nodes with strong correlation, and gets the initial community structure using these pairs of nodes. Lastly, our algorithm detects deterministic community structure by adjusting the fuzzy node and communities that unsatisfied the criteria of community.

The extensive experiments on both real-world and synthetic networks demonstrate the advantages of CAN from three aspects. First of all, the algorithm is very simple. Secondly, our algorithm requires no prior knowledge on the community structure and can obtain the same resulting partition in multiple runs. Lastly, the adjustment of the parameter $\beta$ allows us to uncover communities hierarchically, which could also allow to detect networks with soft communities\cite{zuev2015emergence}.

Finally, we would like to emphasize that some meaningful extensions can be made to CAN in the future. With some improvement, it can be used for uncovering overlapping community.




\begin{thebibliography}{10}
	
	\bibitem{FORTUNATO20161}
	Santo Fortunato and Darko Hric.
	\newblock Community detection in networks: A user guide.
	\newblock {\em Physics Reports}, 659:1 -- 44, 2016.
	\newblock Community detection in networks: A user guide.
	
	\bibitem{newman2012communities}
	Mark~EJ Newman.
	\newblock Communities, modules and large-scale structure in networks.
	\newblock {\em Nature physics}, 8(1):25, 2012.
	
	\bibitem{girvan2002community}
	Michelle Girvan and Mark~EJ Newman.
	\newblock Community structure in social and biological networks.
	\newblock {\em Proceedings of the national academy of sciences},
	99(12):7821--7826, 2002.
	
	\bibitem{PhysRevE.69.026113}
	M.~E.~J. Newman and M.~Girvan.
	\newblock Finding and evaluating community structure in networks.
	\newblock {\em Phys. Rev. E}, 69:026113, Feb 2004.
	
	\bibitem{PhysRevE.90.062805}
	Darko Hric, Richard~K. Darst, and Santo Fortunato.
	\newblock Community detection in networks: Structural communities versus ground
	truth.
	\newblock {\em Phys. Rev. E}, 90:062805, Dec 2014.
	
	\bibitem{PhysRevE.74.036104}
	M.~E.~J. Newman.
	\newblock Finding community structure in networks using the eigenvectors of
	matrices.
	\newblock {\em Phys. Rev. E}, 74:036104, Sep 2006.
	
	\bibitem{Newman8577}
	M.~E.~J. Newman.
	\newblock Modularity and community structure in networks.
	\newblock {\em Proceedings of the National Academy of Sciences},
	103(23):8577--8582, 2006.
	
	\bibitem{PhysRevE.70.066111}
	Aaron Clauset, M.~E.~J. Newman, and Cristopher Moore.
	\newblock Finding community structure in very large networks.
	\newblock {\em Phys. Rev. E}, 70:066111, Dec 2004.
	
	\bibitem{PhysRevE.76.036106}
	Usha~Nandini Raghavan, R\'eka Albert, and Soundar Kumara.
	\newblock Near linear time algorithm to detect community structures in
	large-scale networks.
	\newblock {\em Phys. Rev. E}, 76:036106, Sep 2007.
	
	\bibitem{YOU2016221}
	Tao You, Hui-Min Cheng, Yi-Zi Ning, Ben-Chang Shia, and Zhong-Yuan Zhang.
	\newblock Community detection in complex networks using density-based
	clustering algorithm and manifold learning.
	\newblock {\em Physica A: Statistical Mechanics and its Applications}, 464:221
	-- 230, 2016.
	
	\bibitem{GONG201471}
	Maoguo Gong, Jie Liu, Lijia Ma, Qing Cai, and Licheng Jiao.
	\newblock Novel heuristic density-based method for community detection in
	networks.
	\newblock {\em Physica A: Statistical Mechanics and its Applications}, 403:71
	-- 84, 2014.
	
	\bibitem{Xu:2007:SSC:1281192.1281280}
	Xiaowei Xu, Nurcan Yuruk, Zhidan Feng, and Thomas A.~J. Schweiger.
	\newblock Scan: A structural clustering algorithm for networks.
	\newblock In {\em Proceedings of the 13th ACM SIGKDD International Conference
		on Knowledge Discovery and Data Mining}, KDD '07, pages 824--833, New York,
	NY, USA, 2007. ACM.
	
	\bibitem{pons2005computing}
	Pascal Pons and Matthieu Latapy.
	\newblock Computing communities in large networks using random walks.
	\newblock In {\em International symposium on computer and information
		sciences}, pages 284--293. Springer, 2005.
	
	\bibitem{Shao2015Community}
	Junming Shao, Zhichao Han, Qinli Yang, and Tao Zhou.
	\newblock Community detection based on distance dynamics.
	\newblock In {\em Proceedings of the 21th ACM SIGKDD International Conference
		on Knowledge Discovery and Data Mining}, KDD '15, pages 1075--1084, New York,
	NY, USA, 2015. ACM.
	
	\bibitem{li2012efficient}
	Kun Li, Xiaofeng Gong, Shuguang Guan, and C-H Lai.
	\newblock Efficient algorithm based on neighborhood overlap for community
	identification in complex networks.
	\newblock {\em Physica A: Statistical Mechanics and its Applications},
	391(4):1788--1796, 2012.
	
	\bibitem{WANG20181344}
	Tao Wang, Liyan Yin, and Xiaoxia Wang.
	\newblock A community detection method based on local similarity and degree
	clustering information.
	\newblock {\em Physica A: Statistical Mechanics and its Applications}, 490:1344
	-- 1354, 2018.
	
	\bibitem{vzalik2015maximal}
	Krista~Rizman {\v{Z}}alik.
	\newblock Maximal neighbor similarity reveals real communities in networks.
	\newblock {\em Scientific reports}, 5:18374, 2015.
	
	\bibitem{zhou2009predicting}
	Tao Zhou, Linyuan L{\"u}, and Yi-Cheng Zhang.
	\newblock Predicting missing links via local information.
	\newblock {\em The European Physical Journal B}, 71(4):623--630, 2009.
	
	\bibitem{Salton:1986:IMI:576628}
	Gerard Salton and Michael~J. McGill.
	\newblock {\em Introduction to Modern Information Retrieval}.
	\newblock McGraw-Hill, Inc., New York, NY, USA, 1986.
	
	\bibitem{Pearson1895Proceedings}
	Karl Pearson.
	\newblock Note on regression and inheritance in the case of two parents.
	\newblock {\em Proceedings of the Royal Society of London}, 58:240--242, 1895.
	
	\bibitem{Jaccard1901Etude}
	Paul Jaccard.
	\newblock Etude de la distribution florale dans une portion des alpes et du
	jura.
	\newblock 37:547--579, 01 1901.
	
	\bibitem{quiles2016dynamical}
	Marcos~G Quiles, Elbert~EN Macau, and Nicol{\'a}s Rubido.
	\newblock Dynamical detection of network communities.
	\newblock {\em Scientific reports}, 6:25570, 2016.
	
	\bibitem{Newman2004Finding}
	M.~E.~J. Newman and M.~Girvan.
	\newblock Finding and evaluating community structure in networks.
	\newblock {\em Phys. Rev. E}, 69:026113, Feb 2004.
	
	\bibitem{Danon2005Comparing}
	Leon Danon, Albert Díaz-Guilera, Jordi Duch, and Alex Arenas.
	\newblock Comparing community structure identification.
	\newblock {\em Journal of Statistical Mechanics: Theory and Experiment},
	2005(09):P09008, 2005.
	
	\bibitem{Lancichinetti2008Benchmark}
	Andrea Lancichinetti, Santo Fortunato, and Filippo Radicchi.
	\newblock Benchmark graphs for testing community detection algorithms.
	\newblock {\em Phys. Rev. E}, 78:046110, Oct 2008.
	
	\bibitem{zachary1977information}
	Wayne~W Zachary.
	\newblock An information flow model for conflict and fission in small groups.
	\newblock {\em Journal of anthropological research}, 33(4):452--473, 1977.
	
	\bibitem{LusseauS186emergent}
	David Lusseau.
	\newblock The emergent properties of a dolphin social network.
	\newblock {\em Proceedings of the Royal Society of London B: Biological
		Sciences}, 270(Suppl 2):S186--S188, 2003.
	
	\bibitem{Riskmap}
	Wikipedia.
	\newblock Risk (game).
	\newblock \url{https://en.wikipedia.org/wiki/Risk_(game)#cite_note-1}.
	
	\bibitem{newman2006modularity}
	Mark~EJ Newman.
	\newblock Modularity and community structure in networks.
	\newblock {\em Proceedings of the national academy of sciences},
	103(23):8577--8582, 2006.
	
	\bibitem{knuth1993stanford}
	Donald~Ervin Knuth.
	\newblock {\em The Stanford GraphBase: a platform for combinatorial computing},
	volume~37.
	\newblock Addison-Wesley Reading, 1993.
	
	\bibitem{PhysRevE.70.056122}
	Mari\'an Bogu\~n\'a, Romualdo Pastor-Satorras, Albert D\'{\i}az-Guilera, and
	Alex Arenas.
	\newblock Models of social networks based on social distance attachment.
	\newblock {\em Phys. Rev. E}, 70:056122, Nov 2004.
	
	\bibitem{zuev2015emergence}
	Konstantin Zuev, Mari{\'a}n Bogu{\~n}{\'a}, Ginestra Bianconi, and Dmitri
	Krioukov.
	\newblock Emergence of soft communities from geometric preferential attachment.
	\newblock {\em Scientific reports}, 5:9421, 2015.
	
\end{thebibliography}


\end{document}